\documentclass[10pt,conference]{IEEEtran}

\usepackage[preprint]{waspaa25}

\usepackage{bm} 

\usepackage{amsmath}
\usepackage{graphicx}
\usepackage{hyperref}
\usepackage{lineno}

\DeclareMathOperator{\adashape}{AdaShape}

\usepackage{tikz}
\usetikzlibrary{shapes.geometric, arrows, backgrounds}
\definecolor{squidink}{RGB}{35, 47, 62}
\definecolor{anchor}{RGB}{0, 49, 129}
\definecolor{sky}{RGB}{32, 116, 213}
\definecolor{rind}{RGB}{251, 216, 191}
\definecolor{smile}{RGB}{255, 153, 0}

\tikzstyle{inout} = [rectangle, rounded corners, minimum width=2cm, minimum height=0.7cm, text centered, draw=black, fill=black!10]
\tikzstyle{conv1d} = [rectangle, rounded corners, minimum width=2.5cm, minimum height=0.7cm, text centered, draw=black, fill=orange!30]
\tikzstyle{pool} = [rectangle, rounded corners, minimum width=2.5cm, minimum height=0.7cm, text centered, draw=black, fill=blue!30]
\tikzstyle{tconv1d} = [rectangle, rounded corners, minimum width=2.5cm, minimum height=0.7cm, text centered, draw=black, fill=orange!70]
\tikzstyle{gru} = [rectangle, rounded corners, minimum width=2.5cm, minimum height=0.7cm, text centered, draw=black, fill=red!70]
\tikzstyle{adaconv1d} = [rectangle, rounded corners, minimum width=6cm, minimum height=0.7cm, text centered, draw=black, fill=red!30!orange!30]
\tikzstyle{adacomb1d} = [rectangle, rounded corners, minimum width=2.5cm, minimum height=0.7cm, text centered, draw=black, fill=red!30!orange!60]
\tikzstyle{arrow}=[draw, -latex]
\tikzstyle{adashape} = [rectangle, rounded corners, minimum width=1.8cm, minimum height=0.7cm, text centered, draw=black, fill=sky!60]
\tikzstyle{nonlin} = [rectangle, rounded corners, minimum width=1.8cm, minimum height=0.7cm, text centered, draw=black, fill=sky!30]
\tikzstyle{upsamp} = [rectangle, rounded corners, minimum width=1.5cm, minimum height=0.7cm, text centered, draw=black, fill=black!5]

\title{A lightweight and robust method for blind wideband-to-fullband extension of speech}

\name{Jan B\"uthe, Jean-Marc Valin}
\address{Xiph.org Foundation}

\begin{document}

\maketitle

\begin{abstract}
  Reducing the bandwidth of speech is common practice in resource constrained environments like low-bandwidth speech transmission or low-complexity vocoding. We propose a lightweight and robust method for extending the bandwidth of wideband speech signals that is inspired by classical methods developed in the speech coding context. The resulting model has just $\sim$370~K parameters and a complexity of $\sim$140~MFLOPS (or $\sim$70~MMACS). With a frame size of 10~ms and a lookahead of only 0.27~ms, the model is well-suited for use with common wideband speech codecs. We evaluate the model's robustness by pairing it with the Opus SILK speech codec (1.5 release) and verify in a P.808 DCR listening test that it significantly improves quality from 6 to 12~kb/s. We also demonstrate that Opus 1.5 together with the proposed bandwidth extension at 9~kb/s meets the quality of 3GPP EVS at 9.6~kb/s and that of Opus 1.4 at 18~kb/s showing that the blind bandwidth extension can meet the quality of classical guided bandwidth extensions thus providing a way for backward-compatible quality improvement.
\end{abstract}

\section{Introduction}
Limiting the bandwidth of speech is a common technique for dealing with constrained resources. The most prominent example is speech coding for real-time communication which often uses narrowband codecs (e.g. G.711 \cite{g711}) or wideband codecs (e.g. AMR-WB \cite{amrwb}, Opus SILK \cite{vos_silk}). A second example is neural vocoding in complexity-constrained environments (e.g. LPCNet \cite{lpcnet}) which is used for many applications like text-to-speech synthesis or speech enhancement.

While bandwidth reduction is effective for saving resources and (mostly) maintains speech intelligibility, it does degrade the listening experience and can lead to listener fatigue. Therefore, a blind bandwith extension (BWE) method can have a largely positive impact for billions of listeners every day. However, low complexity is critical for the applications stated above as typical target devices like smartphones can have rather limited compute. Furthermore, robustness is essential since any real-world deployment faces huge variability of input speech signals.

BWE is a well-studied topic and both classical (\cite{makhoul79_bwe, cheng94_bbwe, yasukawa96_tdbwe, yasukawa96_fdbwe, venkatraman15_tdbwe}) and DNN-based (\cite{eskimez_super_resolution, soltanmohammadi_super_resolution, Liu22_vocoding_bwe, mandel23_aero, gomez23_bbwe}) methods have been proposed for this task. While classical methods often have low complexity, they struggle with blind highband estimation and are therefore most effective when provided with side information. DNN-based methods, on the other hand, are much better at high-band modeling, but even dedicated low-complexity algorithms still operate in the range of multiple GFLOPS (e.g. $\sim$ 13~GFLOPS in \cite{soltanmohammadi_super_resolution} or $\sim$7~GFLOPS in \cite{gomez23_bbwe}) which prevents their deployment on mobile devices.

In this paper we seek to overcome this problem by combining the high-band modeling capacity of data-driven, DNN-based methods with the simplicity and low complexity of DSP-based BWE methods. The approach is inspired by classical time-domain bandwidth extension, where a bandwidth-extending operation like non-linear function application or spectral folding is applied to the upsampled signal and combined with time-varying spectral shaping filters. The signal-processing part of the resulting algorithm only consists of classical DSP, i.e. fixed non-linear mapping, fixed and time-varying linear filtering and time-varying sample-wise weighting. The time-varying filters and sample-wise weights in turn are adapted by a small DNN which governs the content and shape of the generated highband signal. The resulting model is trained with a mixture of regression and adversarial losses. It has $\sim$370~K parameters and a computational complexity of $\sim$140~MFLOPS (or $\sim$70~MMACS) which make it suitable for use even on older smartphone devices. Furthermore, since it is built around a low-delay upsampler, it requires only a lookahead of 0.27~ms in addition to a framing delay of 10~ms that may be shared with the wideband speech production system.

To test model robustness, we combine it with the Opus codec (1.5 release) and confirm in a P.808 listening test\footnote{The page \href{https://janpbuethe.github.io/BWEDemo}{https://janpbuethe.github.io/BWEDemo} contains some demo samples. It also includes vocoding examples which were not included in the listening test.} that the BWE model, though only trained on clean speech, provides consistent improvement for all tested bitrates. We furthermore include the superwideband codec EVS \cite{evs} at 9.6~kb/s and Opus 1.4 at 18~kb/s which produces fullband speech in a hybrid coding mode. The test results show that both are statistically tied with the bandwidth-extended Opus 1.5 at 9~kb/s showing that Opus 1.5 with blind bandwidth extension can meet the quality of codecs with classical parametric resp. semi-parametric bandwidth extensions.

A python and C implementation is available at \href{https://gitlab.xiph.org/xiph/opus/-/tree/waspaa_2025_bwe}{https://gitlab.xiph.org/xiph/opus/-/tree/waspaa\_2025\_bwe} (BBWENet).

\section{Model Description}
\begin{figure}
\center
\resizebox{\columnwidth}{!}{
\begin{tikzpicture}[node distance=1.2cm]

\def\signalx{7}

\node (features) [inout] at (0, 0) {Features};
\node (fconv1) [conv1d] at (0, -1.5) {Conv(k=3,s=1)};
\node (fconv2) [conv1d] at (0, -3) {Conv(k=3,s=1)};
\node (ftconv) [tconv1d] at (0, -4.5) {TConv(k=2,s=2)};
\node (fgru) [gru] at (0, -6) {GRU};
\node (phi) at (0, -7.5) {\small$\varphi_n$};

\def\deltay{2}
\def\signalyI{0}
\node (sigin)  at (\signalx, \signalyI) {$y_{16}(t)$};

\def\signalyII{\signalyI - 1.5}
\node (adaconv1) [adaconv1d] at (\signalx, \signalyII) {AdaConv(k=15)};

\def\signalyIII{\signalyII-1.5}
\node (up11) [upsamp] at (\signalx-2, \signalyIII) {$2\uparrow$};
\node (up12) [upsamp] at (\signalx, \signalyIII) {$2\uparrow$};
\node (up13) [upsamp] at (\signalx+2, \signalyIII) {$2\uparrow$};

\def\signalyIV{\signalyIII-1.5}
\node (adashape1) [adashape] at (\signalx, \signalyIV) {AdaShape};
\node (nonlin1) [nonlin] at (\signalx + 2, \signalyIV) {NonLin};

\def\signalyV{\signalyIV-1.5}
\node (adaconv2) [adaconv1d] at (\signalx, \signalyV) {AdaConv(k=25)};

\def\signalyVI{\signalyV-3}
\node (up21) [upsamp] at (\signalx-2, \signalyVI) {$1.5\uparrow$};
\node (up22) [upsamp] at (\signalx, \signalyVI) {$1.5\uparrow$};
\node (up23) [upsamp] at (\signalx+2, \signalyVI) {$1.5\uparrow$};

\node (y32) at (\signalx-2, \signalyV - 1.5) {$y_{32}(t)$};

\def\signalyVII{\signalyVI-1.5}
\node (adashape2) [adashape] at (\signalx, \signalyVII) {AdaShape};
\node (nonlin2) [nonlin] at (\signalx + 2, \signalyVII) {NonLin};

\def\signalyVIII{\signalyVII-1.5}
\node (adaconv3) [adaconv1d] at (\signalx, \signalyVIII) {AdaConv(k=15)};

\node (sigout) at (\signalx, \signalyVIII - 1.5) {$y_{48}(t)$};

\draw [arrow, dashed] (features) -- (fconv1);
\draw [arrow, dashed] (fconv1) -- (fconv2);
\draw [arrow, dashed] (fconv2) -- (ftconv);
\draw [arrow, dashed] (ftconv) -- (fgru);

\draw [arrow, dashed] (sigin) -- (features);

\def\phix{2.5}
\draw [arrow, dashed] (fgru) -- (phi);
\draw [arrow, dashed] (phi) -- (\phix, -7.5) -- (\phix, \signalyII) -- (adaconv1);
\draw [arrow, dashed] (\phix, \signalyIV) -- (adashape1);
\draw [arrow, dashed] (\phix, \signalyV) -- (adaconv2);
\draw [arrow, dashed] (\phix, \signalyVII) -- (adashape2);
\draw [arrow, dashed] (\phix, -7.5) -- (\phix, \signalyVIII) -- (adaconv3);


\draw [arrow] (sigin) -- (adaconv1);

\draw [arrow] (\signalx-2, \signalyII - 0.35) -- (up11);
\draw [arrow] (adaconv1) -- (up12);
\draw [arrow] (\signalx+2, \signalyII - 0.35) -- (up13);

\draw [arrow] (up12) -- (adashape1);
\draw [arrow] (up13) -- (nonlin1);

\draw [arrow] (up11) -- (\signalx - 2, \signalyV + 0.35);
\draw [arrow] (adashape1) -- (\signalx, \signalyV + 0.35);
\draw [arrow] (nonlin1) -- (\signalx + 2, \signalyV + 0.35);

\draw (\signalx-2, \signalyV - 0.35) -- (y32);
\draw [arrow]  (y32) -- (up21);
\draw [arrow] (adaconv2) -- (up22);
\draw [arrow] (\signalx+2, \signalyV - 0.35) -- (up23);

\draw [arrow] (up22) -- (adashape2);
\draw [arrow] (up23) -- (nonlin2);

\draw [arrow] (up21) -- (\signalx - 2, \signalyVIII + 0.35);
\draw [arrow] (adashape2) -- (\signalx, \signalyVIII + 0.35);
\draw [arrow] (nonlin2) -- (\signalx + 2, \signalyVIII + 0.35);

\draw [arrow] (adaconv3) -- (sigout);


\begin{pgfonlayer}{background}
\filldraw [line width=4mm, join=round, black!10]
(fconv1.north -| fconv1.east) rectangle (fgru.south -| fgru.west)
(adaconv1.north -| adaconv1.east) rectangle (adaconv3.south -| adaconv3.west);
\end{pgfonlayer}{background}

\end{tikzpicture}
}
\caption{High-level overview of blind bandwidth-extension model. The feature encoder on the left side calculates a sequence of latent feature vectors from a sequence of 72-dimensional feature vector containing spectral information. The latent feature vectors are then used by AdaConv modules to steer pre- and post-filtering and by the AdaShape module by adaptively extending the bandwidths of the input signals.}\label{figure:model}
\end{figure}
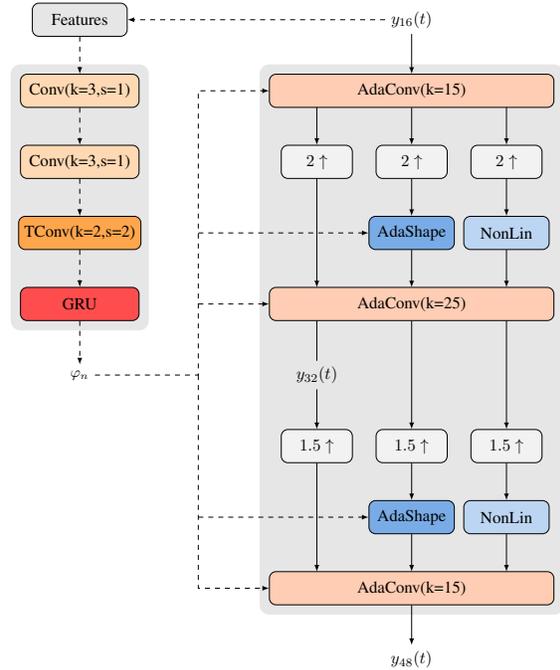

We propose a model based on the classical approach of pre-filtering, upsampling, bandwidth extension and post-filtering \cite{makhoul79_bwe}. A high-level overview of the model is given in Figure \ref{figure:model}.

Adaptive pre- and post-filtering of the signal is implemented using the AdaConv module proposed in \cite{buethe23_lace} and extended to multiple input and output channels in \cite{buethe24_nolace}. AdaConv is similar to regular Conv1d layers but the weights are adapted at a fixed rate (200 Hz in this case) based on a latent feature vector provided by the feature encoder depicted on the left side of Figure \ref{figure:model}.

For upsampling, the model leverages the libopus 16-to-48-kHz upsampler\footnote{\href{https://gitlab.xiph.org/xiph/opus}{https://gitlab.xiph.org/xiph/opus}} which operates in two stages: in a first stage the signal is upsampled by a factor two using IIR filters and in a second stage a 1.5x interpolation with short FIR filters is performed. The upsampler has low complexity and induces a low delay of 13 samples at 48~kHz which is also the total delay of the signal path on the right hand side of Figure \ref{figure:model}. IIR filters are approximated by sufficiently long FIR filters for training.

For the actual bandwidth extension, the model uses a hybrid approach that utilizes two common methods for time-domain bandwidth extension. The first one is the application of a non-linear function, a method that generates a consistent harmonic extension for quasi-periodic lowbands (voiced speech). We deviate from the usual choice of non-linearity (absolute value or rectified linear unit) and use a non-linearity instead that extends the signal more agressively \eqref{eq:jm-nonlin}. It is designed to approximately preserve the scale of the signal and to induce a similar amount of distortion regardless of the scale of the input signal. 

\begin{equation}\label{eq:jm-nonlin}
    f(x) = x \sin(\log\mid x\mid)
\end{equation}

The second extension method is motivated by spectral folding as proposed in \cite{makhoul79_bwe}. We use the term spectral folding in a broader sense of multiplying the signal with a locally periodic sequence of non-negative weights which we implement using the AdaShape module proposed in \cite{buethe24_nolace}. It multiplies the input signal with a sequence of weights calculated from a sequence of latent feature vectors.
\begin{equation}
\adashape(x(\cdot), \phi(\cdot))(n) = \alpha(n, \phi(\cdot), x(\cdot)) \cdot x(n).
\end{equation}

Folding, especially when combined with spectral flattening as pre-filtering, provides an effective way for extending unvoiced signal parts. In principle, this module is also capable of extending the signal by sharpening pulses in voiced signal parts. However, an a-posteriori analysis (Figure \ref{fig:example}) shows the model mostly uses folding for extending unvoiced signal parts and the non-linear function extrapolation for extending voiced signal parts.

The input features on the left side of Figure \ref{figure:model} are designed to provide the model with basic signal properties like spectral envelope, pitch and voicing while being simple to compute. We chose to use a Hanning-window STFT with a window size of 20~ms and a hop size of 10~ms from which we compute a 32-band ERB-scale log magnitude spectrogram and complex phase difference information for the first 40 STFT bins in the form
\begin{equation}
\Delta\Phi(k, n) = \frac{X(n, k) \, X^*(n-1, k)}{\mid X(n, k) \, X^*(n-1, k) \mid},
\end{equation}
where $n$ denotes the frame index, $k$ denotes the frequency index, and $^*$ denotes complex conjugation. The complex phase differences are included as features since they proved sufficient for high-accuracy pitch estimation \cite{subramani24_npitch}. The feature encoder upsamples the 72-dimensional feature vector from 100 to 200~Hz and includes a GRU for accumulating context.

\section{Training}
\subsection{Strategy and Data}
Bandwidth extension, viewed as an inverse problem, is generally ill posed since even for the same source signal, the actual recording will depend on environmental factors such as acoustic environment or recording device. It is possible to a certain extent to fit a model to the characteristics of a specific homogeneous dataset, but we observed that this can lead to poor generalisation. This is in line with the findings in \cite{huber2004robust} which in particular highlights the critical impact of microphone channels. The training procedure for the proposed model therefore prioritizes plausibility and robustness over correctness.

To achieve plausibility, we follow the standard approach and use an adversarial loss to bring the distribution of the extended signal closer to the distribution of a native fullband signal \cite{eskimez_super_resolution}. We do, however, use a family of frequency-domain discriminators instead of the commonly used multi-scale and multi-period time-domain discriminators as we found this to lead to faster convergence and higher quality.

We train the model on a mixture of multiple high-quality TTS datasets \cite{demirsahin-etal-2020-open, kjartansson-etal-2020-open, kjartansson-etal-tts-sltu2018, guevara-rukoz-etal-2020-crowdsourcing, he-etal-2020-open, oo-etal-2020-burmese, van-niekerk-etal-2017, gutkin-et-al-yoruba2020, bakhturina21_interspeech} containing more than 900 speakers in 34 languages and dialects. Since some of the datasets contained upsampled wideband and superwideband recordings, we filter out items with very little energy high frequency range. Furthermore, we apply the following data augmentation procedures to increase model robustness to unseen conditions:
\begin{enumerate}
\item we apply a random eq filter that is constant above 4~kHz to 40\% of the training clips\label{i:random-eq}
\item we add stationary wideband noise with random gain to 20\% of training clips\label{i:random-noise}
\item  we apply a random RIR from the Aachen Impulse Response Database\footnote{\href{https://www.openslr.org/20}{https://www.openslr.org/20}} to 20\% of training clips\label{i:random-rir}
\item we add a random DC offset to 10\% of training clips\label{i:random-dc}
\end{enumerate}
Finally, we filter the 48~kHz target clips with a 20~kHz lowpass filter to remove ambiguity from mixing 44.1 and 48~kHz speech samples.

Augmentation \ref{i:random-eq}) prevents the model from relying too much on low frequencies, which should be irrelevant for extending the wideband signal. Augmentation \ref{i:random-noise}) teaches the model not to extend noise but only the speech from the baseband. This is an intentional design decision and it stems from the observation, that adding a bandwidth extension to a signal of poor quality can reduce perceived quality as the resulting signal sounds noisier than the bandlimited input signal. Augmentations \ref{i:random-noise}) and \ref{i:random-rir}), when both carried out on the same item, are applied in random order.

The model input is derived from the augmented 48~kHz signals by applying a random lowpass filter with cutoff between 7.5 and 8~kHz and varying slope. Furthermore, the target signal is delayed by 13 samples to compensate for the resampling delay in the signal path.

\subsection{Losses and Training}
Training is split into a pre-training phase, using only regression losses, and an adversarial training phase using both a discriminator loss and a regression loss for regularization.

We use three regression losses, the STFT-based envelope matching loss and the spectral fine structure loss $\mathcal{L}_{env}$ and $\mathcal{L}_{spec}$ from \cite{buethe23_lace}, which are averaged over multiple STFT resolutions with window sizes $3\cdot 2^n$ for $6 \leq n \leq 11$, and a time-domain $L^2$ loss $\mathcal{L}_{tdlp}$ on the low frequency range to enforce lowband reconstruction. The lowpass filter is a 15-tap, zero-phase filter with a cutoff frequency of 4~kHz and a gentle rolloff. The total regression loss for pretraining is given by
\begin{equation}
\mathcal{L}_{pre} = \frac{1}{13} \mathcal{L}_{env} + \frac{2}{13} \mathcal{L}_{spec} + \frac{10}{13} \mathcal{L}_{tdlp}
\end{equation}
Pretraining is carried out on 1s-segments using the Adam optimizer with a batch size of 256, an initial learning rate of $5\times 10^{-4}$ and a weight decay factor of $2.5\times 10^{-5}$ for 50 epochs.

For adversarial training, we use a modification of the discriminator architecture in \cite{buethe24_nolace}, a multi-layer 2D-convolutional model on log-magnitude spectrograms including frequency-positional embeddings to allows the discriminator to take the frequency range into account. The main changes are increasing the STFT sizes by a factor 3 to compensate for the sampling rate increase, increasing the kernel size along the frequency axis from from $3\times 3$ to $7\times 3$ to maintain the frequency widths of receptive fields, and limiting the maximal channel number to 64 to compensate for the parameter increase caused by the wider covolution kernels. Otherwise, adversarial training is identical to \cite{buethe24_nolace} with $\mathcal{L}_{reg} = 0.6 \mathcal{L}_{pre}$. Adversarial training is carried out on $0.9$-second segments with a batch size of 64 using the Adam optimizer with a constant learning rate of $10^{-4}$ for 40 epochs.

\section{Evaluation}
\subsection{Subjective Evaluation}\label{ss:subjective}

\begin{figure}
    \centering
    \includegraphics[width=\columnwidth]{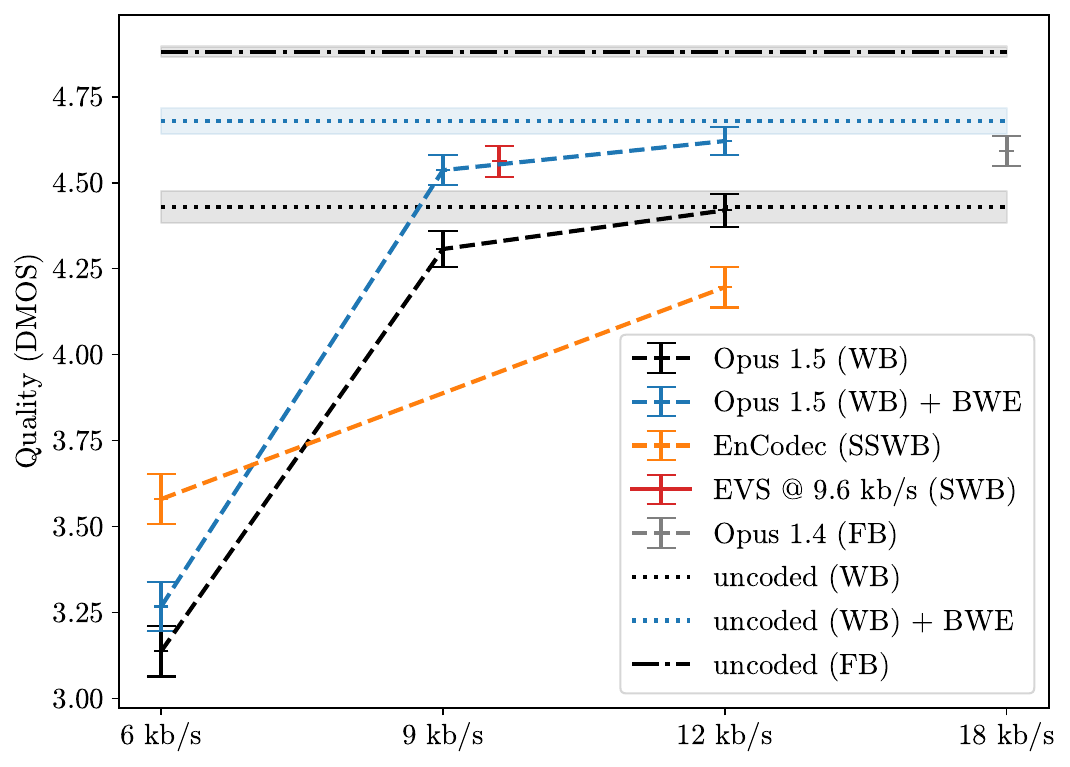}
    \caption{Results of the P.808 DCR listening on samples of the EARS dataset. Bandwidth-extended signals are labeled with '+ BWE'. All bandwidth-extended signals show significant improvement over their wideband source signals ($p=0.95$). Furthermore, extended Opus 1.5 can match the quality of higher bandwidth codecs that use classical, guided methods for coding above-wideband content.}
    \label{fig:mos}
\end{figure}

\begin{figure*}[h]
    \centering
    \includegraphics[width=0.75\textwidth]{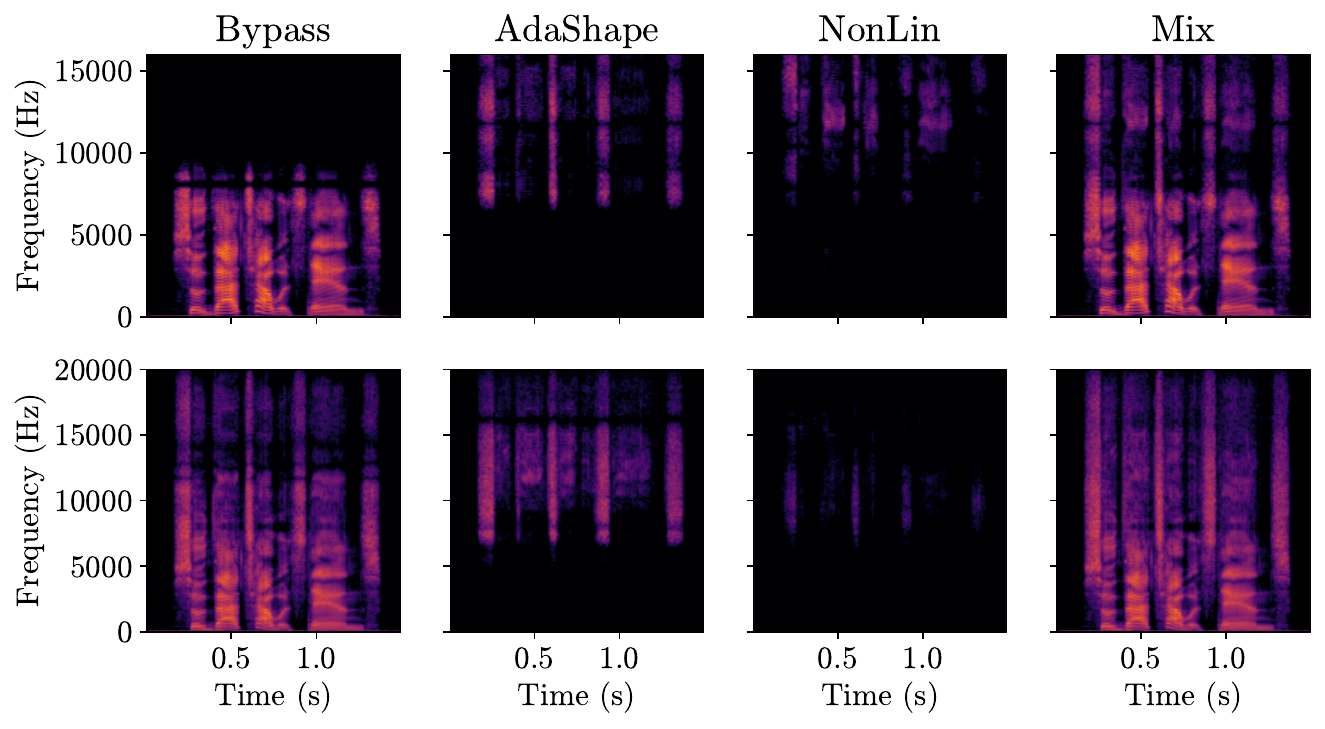}
    \caption{A decomposition of the signals $y_{32}(t)$ (top) and $y_{48}(t)$ (bottom) as sum of the bypass, AdaShape and NonLin contributions. Spectrograms show the model mainly uses AdaShape for extending unvoiced speech and NonLin for extending voiced speech.}
    \label{fig:example}
\end{figure*}

To test model performance and robustness, we carried out a multi-bitrate speech coding test using the open source implementation \cite{naderi2020} of the P.808 DCR methodology. We sampled three random sentence pairs per speaker from the EARS dataset \cite{richter24_ears} (regular speech category) as test material and applied loudness normalization. Note that no condition was trained on the EARS dataset. We tested the proposed bandwidth extension both for Opus 1.5 with decoder complexity 10 (which means NoLACE-enhancement \cite{buethe24_nolace} will be applied) and for clean speech input and report significant improvement for all conditions. In particular, the improvement of Opus 1.5 at 6~kb/s is remarkable since the baseband already exhibits very audible distortion.

As comparison points we included the 3GPP EVS codec at 9.6~kb/s\footnote{EVS codes frequencies up to $14.4$ kHz at this bitrate} and the results show that Opus with lowband enhancement and blind bandwidth extension matches the quality even at the same bitrate. This indirectly gives a comparison of the proposed blind BWE method to a guided, parametric BWE. Furthermore, we included EnCodec at 6 and 12~kb/s to test the hybrid approach (mix of DNN and DSP) against the full end-to-end neural coding approach. While EnCodec delivers better quality at 6~kb/s, even Opus 1.5 wideband at 9~kb/s already significantly outperforms EnCodec at 12~kb/s. This likely highlights a lack of robustness to out-of-domain data, as the quality reported in \cite{defossez_encodec} is much higher. In particular, a pre-test also included the neural codec AudioDec, which exhibited severe quality degradation on this dataset and was therefore excluded from the final test (examples are included in the demo page). These results suggest, that the hybrid approach may be more robust than the end-to-end approach and they certainly suggest that robustness must be carefully evaluated when considering to deploy an end-to-end codec.

Finally, we included Opus 1.4 at 18~kb/s as a reference point to evaluate the combined effect of lowband enhancement and highband extension. Opus 1.5 at 9~kb/s is statistically tied to Opus 1.4 at 18~kb/s, with equivalent quality likely achieved around 10~kb/s, which results in a bitrate reduction of 45 to 50\%. Additionally, this comparison provides an indirect evaluation of the blind BWE against an Opus-coded highband in hybrid mode.

The evaluation on clean, uncoded speech shows, that while the proposed bandwidth extension significantly improves quality in a direct comparison to the fullband reference signal it is still distinguishable from the original.

\subsection{Model Inspection}

Due to the simple architecture of the signal-processing part of the model, it is straightforward to inspect the contribution of the individual modules to the final bandwidth extension. In particular, since the mixing of the extension layers is linear, the second bypass signal $y_{32}(t)$ and the output signal $y_{48}(t)$ can be decomposed as a sum of signals stemming from the previous bypass channel and the output channels of the AdaShape and NonLin modules.

Spectrograms of these contributions are displayed in Figure \ref{fig:example}. It shows that in the first stage, unvoiced signal parts are primarily extended by the AdaShape module while voiced signal parts are mostly extended by the non-linearity. In the second stage, the SWB to FB extension is primarily constructed from the AdaShape output and from imaging remaining from the short FIR interpolation filters.

This analysis shows the usefulness of the dual approach of combining these two extension. It also suggests the second NonLin module could likely be omitted without loss of quality which would result in a small complexity saving. While no formal listening test was carried out on this matter, the authors observed by informal (blind) listening that omitting either the non-linearity or the AdaShape module from the model results in audible degradation of the extended signal.

\section{Conclusion}
We proposed a lightweight method for wideband-to-fullband extension of speech and demonstrated that it improves wideband speech from multiple sources. In particular, the proposed model enables fullband speech coding with the Opus codec down to very low bitrates without adding additional delay. Our analysis furthermore shows, that the model adopts a dual approach for extending speech, using folding for unvoiced and non-linear extension for voiced speech parts. This insight can potentially help improve classical BWEs in the future. A next step will be to adopt the model to also extend music where first tests showed promising results.
 

\bibliographystyle{IEEEtran}
\bibliography{audio_full}







\end{document}